\documentclass[aps,prl,twocolumn,showpacs,superscriptaddress]{revtex4}

\usepackage{graphicx,amsmath,verbatim}

\pdfoutput=1
\newcommand{\be}{\begin{equation}}
\newcommand{\ee}{\end{equation}}
\newcommand{\deff}{d_\mathrm{eff}}
\newcommand{\F}{\mathcal{F}}

\begin{document}

\title{Scaling laws of the cavity enhancement for nitrogen-vacancy centers in diamond}

\author{Hanno Kaupp}
\affiliation{Fakult{\"a}t f{\"u}r Physik, Ludwig-Maximilians-Universit{\"a}t, Schellingstra{\ss}e~4, 80799~M{\"u}nchen, Germany}
\affiliation{Max-Planck-Institut f{\"u}r Quantenoptik,  Hans-Kopfermann-Str.~1, 85748~Garching, Germany}
\author{Christian Deutsch}
\affiliation{Menlo Systems GmbH, 82152 Martinsried, Germany}
\author{Huan-Cheng Chang}
\affiliation{Institute of Atomic and Molecular Sciences, Academia Sinica, Taipei 106, Taiwan}
\author{Jakob Reichel}
\affiliation{Laboratoire Kastler Brossel, ENS/UPMC-Paris 6/CNRS, 24
  rue Lhomond, F-75005 Paris, France}
\author{Theodor W. H{\"a}nsch}
\affiliation{Fakult{\"a}t f{\"u}r Physik, Ludwig-Maximilians-Universit{\"a}t, Schellingstra{\ss}e~4, 80799~M{\"u}nchen, Germany}
\affiliation{Max-Planck-Institut f{\"u}r Quantenoptik,  Hans-Kopfermann-Str.~1, 85748~Garching, Germany}
\author{David Hunger}
\email[To whom correspondence should be addressed. E-mail: ]{david.hunger@physik.lmu.de}
\affiliation{Fakult{\"a}t f{\"u}r Physik, Ludwig-Maximilians-Universit{\"a}t, Schellingstra{\ss}e~4, 80799~M{\"u}nchen, Germany}
\affiliation{Max-Planck-Institut f{\"u}r Quantenoptik,  Hans-Kopfermann-Str.~1, 85748~Garching, Germany}

\date{\today}

\begin{abstract}
We employ a fiber-based optical microcavity with high finesse to study the enhancement of phonon sideband
fluorescence of nitrogen-vacancy centers in nanodiamonds. Harnessing the full tunability and open access of
the resonator, we explicitly demonstrate the scaling laws of the Purcell enhancement by varying both the mode
volume and the quality factor over a large range. While changes in the emission lifetime remain small in the regime
of a broadband emitter, we observe an increase of the emission spectral density by up to a factor of 300. This gives
a direct measure of the Purcell factor that could be achieved with this resonator and an emitter whose linewidth
is narrower than the cavity linewidth. Our results show a method for the realization of wavelength-tunable
narrow-band single-photon sources and demonstrate a system that has the potential to reach the strong-coupling regime.
\end{abstract}

\pacs{42.50.Pq, 42.50.Ct, 42.81.Qb, 61.72.jn, 81.05.ug}

\maketitle

The negatively charged nitrogen-vacancy (NV) center in diamond stands out due to stable single photon emission \cite{Gruber97,Kurtsiefer00} and exceptional spin coherence of electronic ground-state levels \cite{Balasubramanian09}. Together with spin-selective optical transitions, the NV center represents a particularly promising solid-state quantum bit with direct optical access \cite{Gruber97,Balasubramanian09,Maurer12}. An important milestone yet to achieve is the realization of an efficient optical interface for this emitter. Ideally it would open the way for deterministic single photon sources \cite{Su08}, non-destructive spin state detection \cite{Young09,Robledo11}, efficient spin-photon entanglement \cite{Togan10}, and quantum coherent optical manipulation on the single photon level.  A variety of optical microcavities has been studied in this respect, including photonic crystal cavities \cite{Wolters10,Englund10,Sar11,Riedrich12,Faraon12}, microring resonators \cite{Faraon11,Hausmann12}, microdiscs \cite{Barclay09,Barclay09b}, Fabry-Perot cavities \cite{Dumeige11}, and plasmonic nanoresonators \cite{Leon12,Choy11}. Considerable success has already been achieved, e.g. by demonstrating Purcell enhancement of the zero phonon line (ZPL) at cryogenic temperature, where $70\%$ of the emission was channeled into a single cavity mode \cite{Faraon11}. The broad emission spectrum and the small weight of the ZPL call for particularly large Purcell factors to achieve efficient cavity coupling. Furthermore, several issues remain problematic for the techniques used so far, which all rely on solid state cavities with limited flexibility to access the cavity mode and to tune the resonance frequency.

In this work we study the cavity enhancement of the phonon sideband emission of small ensembles of NV centers in individual nanodiamonds by means of a fiber based Fabry-Perot microcavity \cite{Hunger10b}. The cavity design relies on concave surface profiles machined on the end faces of optical fibers using CO$_2$ laser processing \cite{Hunger12}. 
The technique provides excellent surface quality to support ultra-low loss mirror coatings, and offers microscopic cavity mode volumes to achieve large Purcell factors. The open resonator geometry permits us to easily access the cavity field maximum, to study various emitters with one and the same cavity, and to fully tune the cavity resonance. These features have been beneficial for recent experiments with ultracold atoms \cite{Colombe07}, trapped ions \cite{Steiner13}, molecules \cite{Toninelli10}, quantum dots \cite{Muller09,Barbour11,Di12,MiguelSanchez13}, and NV centers in diamond \cite{Albrecht13}.
Here we use these properties 
to explicitly demonstrate the scaling of the fluorescence enhancement for NV centers with the cavity parameters. Using direct comparison of free space and cavity enhanced emission, we determine the ideal and effective Purcell factors as a function of the mode volume and the quality factor. While narrow cavity resonances couple only to a small fraction of the broad emission spectrum, we observe an up to 300-fold enhancement of the emission spectral density, giving a direct measure of the ideal Purcell factor of the coupled system. Furthermore, we find that we can maintain a cavity finesse $\F = 30000$ in the presence of rather large nanocrystals.

\begin{figure}[t]
\includegraphics[width=0.45\textwidth]{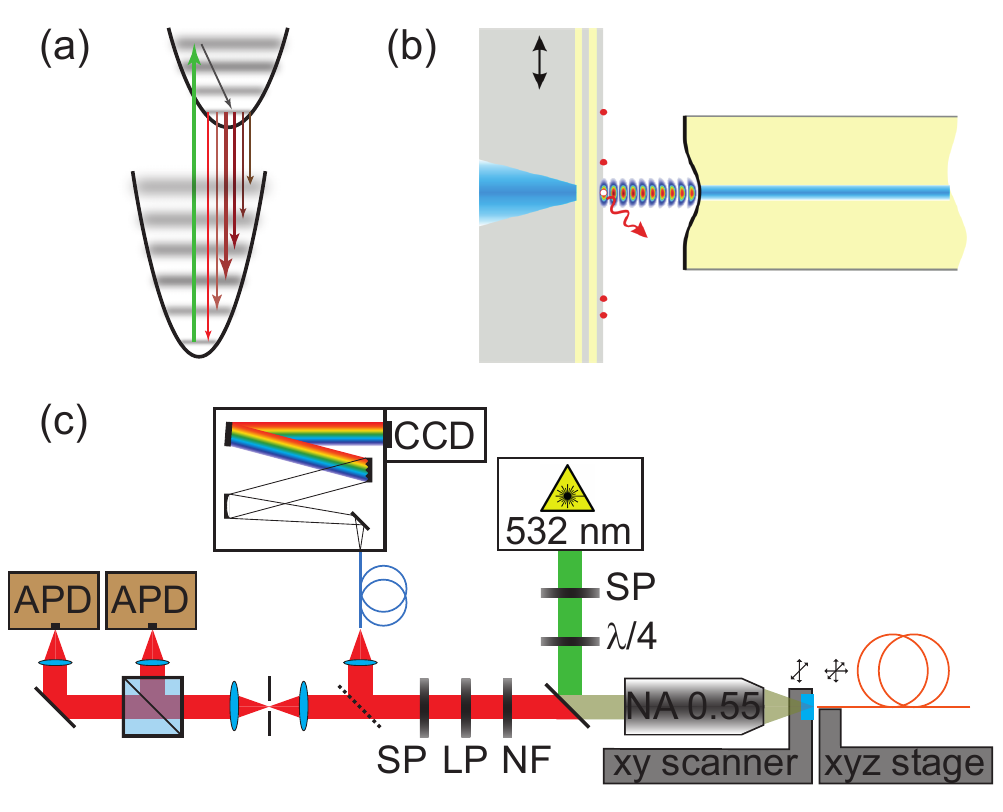}
\caption{\label{fig:Setup} (Color online) (a) Level scheme of phonon assisted optical transitions in the NV optical cycle. (b) Schematic of the cavity. (c) Experimental setup combining a confocal microscope and a tunable microcavity (LP long pass, SP short pass, NF notch filter, $\lambda/4$ quarter wave plate).
}
\end{figure}

\textit{Setup.} In our experiments we use a cavity which consists of a mirror on a concave end face of a single mode fiber and a macroscopic plane mirror, schematically shown in Fig.~\ref{fig:Setup}(b). The effective radius of curvature in the center of the profile machined on the fiber is $r=100~\mu$m. Together, the two mirrors define a fully accessible microscopic cavity with a mode waist $w_0=2.2~\mu$m and an effective cavity length as small as $\deff=4.3~\mu$m, which includes the penetration of the cavity mode into the dielectric mirrors. The smallest mode volume we currently achieve in this way is $V=16~\mu$m$^3$. The coatings are designed for highest reflectivity at 780~nm and are realized by multiple layer pairs of Ta$_2$O$_5$ and SiO$_2$. The finesse reaches a value of $F=30000$ at 780~nm for small cavity length. It decreases continuously towards $\mathcal{F}=20000$ at $\deff=39~\mu$m and drops faster for larger distances, which we attribute to diffraction loss due to the finite mirror size. The largest quality factor we thus obtain is $Q_0=2\times10^6$ for $\deff = 39~\mu$m.

\begin{figure*}[ht]
\includegraphics[width=0.75\textwidth]{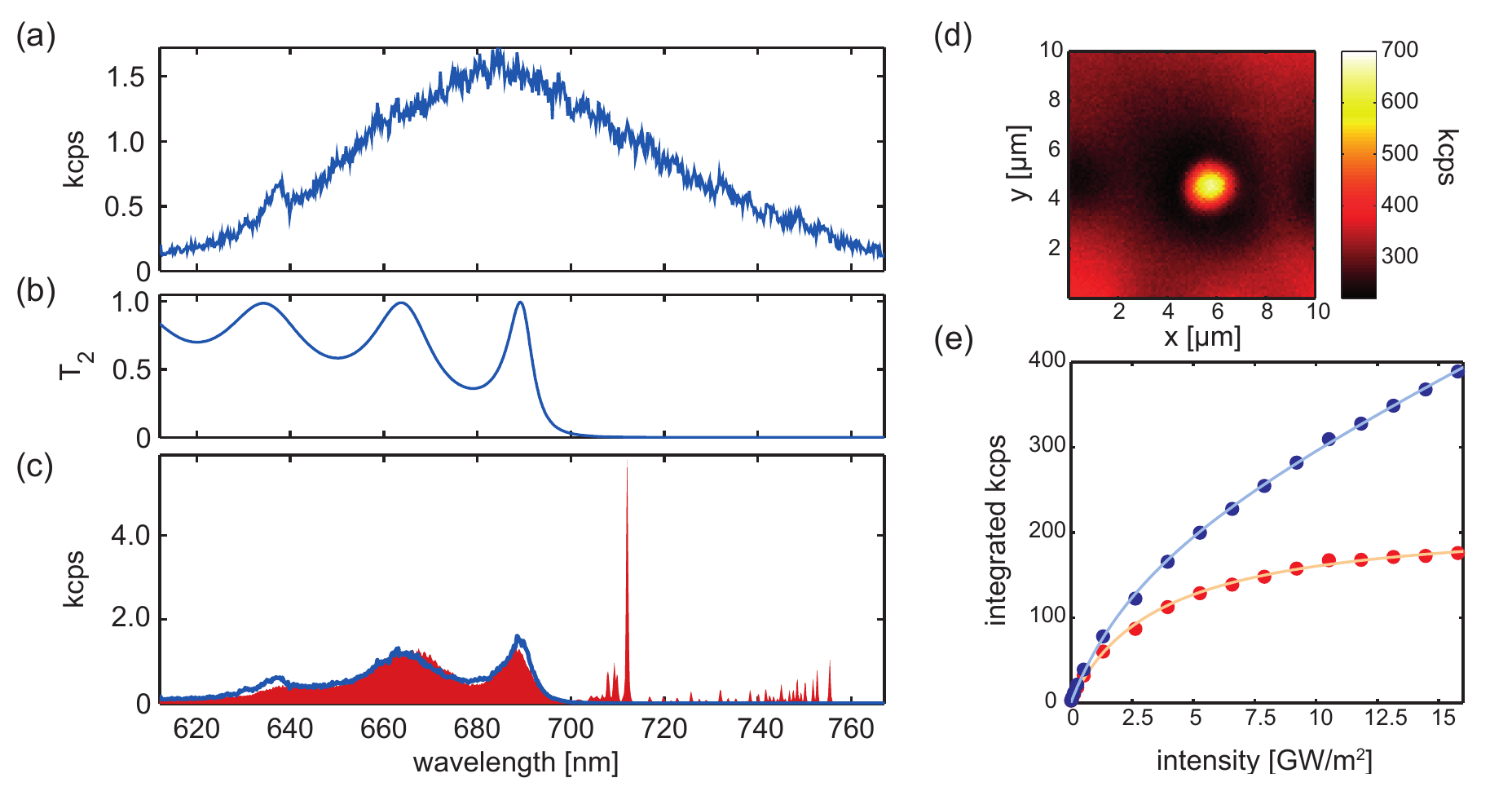}
\caption{\label{fig:Results} (Color online) (a) Free space spectrum of a 100~nm diamond on a suprasil substrate (in units kilocounts per second). (b) Calculated transmission spectrum of the macroscopic mirror. (c) Emission spectrum of a FND placed inside the cavity. The transmissive part can be explained by the free space spectrum and the mirror transmission (blue line). Within the mirror stop band, cavity enhanced emission into narrow resonances is observed. (d) Confocal scan with $I = 0.5~I_\mathrm{sat}$ showing a single nanodiamond, mirror background, and the effect of bleaching. (e) Saturation of free space NV emission (blue datapoints). For comparison we show the same measurement with linear background subtracted (red datapoints). Solid lines are fits to a saturation model.
}
\end{figure*}

We introduce nanodiamonds containing NV centers into the cavity by directly spin coating an aqueous colloidal solution on the large mirror with low enough concentration ($<0.01~\mu$m$^{-2}$) to isolate single nanocrystals with the cavity mode. For optimal coupling between the emitter and the cavity, the mirror is terminated with a $\lambda/4$ layer of SiO$_2$ to shift the electric field maximum of the cavity mode slightly above the mirror surface. We employ two different samples of fluorescent nanodiamonds (FNDs) with size distributions peaking at 30 and 100~nm diameter. They were irradiated with He ions to increase the concentration of NV centers \cite{Mohan10,Yu05}. Surface treatment with acids and thermal oxygen etching removed graphitic surface shells and other contaminants. The resulting surfaces are predominantly oxygen terminated, which favors the negative charge state of the NV center \cite{Rondin10,Hauf11}.
The emission spectrum of the NV center shows a zero phonon line at 637~nm with a room temperature linewidth of $\sim 2~$THz \cite{Fu09}, and a broad phonon sideband resulting from coupling to a phonon continuum with a distinct coupling strength maximum around a phonon frequency of $16$~THz \cite{Davies74}. Transitions involving several phonons constitute the sideband fluorescence as schematically shown in Figure \ref{fig:Setup}(a). The strength of the integrated $k$-th sideband can be estimated by the Frank-Condon factors for displaced harmonic oscillator states, $\zeta_k = e^{-D}D^k/k!$, with a Huang-Rhys factor $D=3.2$ found for NVs in bulk. In our experiments, we observe spectra peaking at a central wavelength of 690~nm with a full width at half maximum of $\sim 70$~nm [see Fig.~\ref{fig:Results}(a)].

With our present mirror coatings, the NV emission spectrum is divided into two parts: For wavelengths shorter than $690~$nm, the mirrors are transmissive and the color centers emit approximately under free space conditions. For longer wavelengths, the mirrors become increasingly reflective and cavity enhanced emission can be observed. This enables us to observe free space and cavity enhanced emission within one measurement, such that the cavity enhancement can be quantified from a direct comparison. Furthermore, it strongly facilitates the search for NV centers in the cavity, since a large fraction of their emission is transmitted. High transparency at the excitation wavelength provides nearly constant excitation conditions when varying the cavity length. The transmission of the plane mirror is chosen to be a factor 2 larger than that of the fiber mirror to optimize outcoupling efficiency into the detection channel used in the experiments. Finally, the strong change of mirror reflectivity across the fluorescence spectrum permits us to study the cavity enhancement as a function of the quality factor.

The cavity setup is combined with a homebuilt confocal microscope as shown in Fig.~\ref{fig:Setup}(c). We use an excitation laser at 532~nm (Cobolt Samba) which is coupled to a singlemode fiber and short pass filtered (Thorlabs FES0550) to remove fiber fluorescence. The light is focused with a long working distance objective (Mitutoyo, NA$=0.55$) through the macroscopic mirror (diameter 0.5 in.) onto the surface supporting the FNDs. The mirror is mounted on a nanopositioning stage (PI 541.2 SL) to perform raster scans and to maximize the overlap of a nanodiamond with the cavity mode. The fluorescence is collected via the same objective, separated from the excitation light with a dichroic mirror (Thorlabs DMLP567), and spectrally filtered (Thorlabs FEL0600, FES0850, NF533-17). Thereafter, the fluorescence is either spatially filtered with a pinhole and led to a Hanbury Brown-Twiss setup with two avalanche photo diodes (Laser Components Count) behind a 50/50 beamsplitter, or coupled via a multimode fiber (Thorlabs HPSC25) to a grating spectrometer (Princeton Instruments Acton SP2500) with a CCD camera (Andor iKon-M).
The cavity fiber is mounted on a three axis micropositioning mount. It can be withdrawn from the macroscopic mirror by some millimeters in order to use the setup as a conventional confocal microscope.

\textit{Characterization.} In a first step we characterize FNDs on Suprasil substrates replacing the macroscopic mirror with the fiber withdrawn. We confirm the presence of negatively charged NV centers by assessing fluorescence spectra, which show no significant contribution of NV$^0$. We find that an appreciable fraction ($>10\%$) of the fluorescent 30~nm diamonds contain single NVs by observing antibunching in the second order autocorrelation function of the photoluminescence. Comparing the average single emitter count rate with the fluorescence yield of 100-nm FNDs, we find that the latter host 30-300 NV centers per nanocrystal.

Next, we study FNDs prepared on a cavity mirror and with the fiber aligned to form a cavity. We perform confocal scans with the mirror surface facing the fiber and detect the partial transmission through the mirror as shown in Fig.~\ref{fig:Results}(d). We observe strong background fluorescence within the detection spectral band, originating from fluorescence generated in the mirror and the fiber, as well as from Raman scattering in the fiber. To obtain signals much stronger than the mirror background, we use 100~nm nanodiamonds containing ensembles of NV centers. With a coating that prevents excitation light from entering the fiber, excessive background is avoidable and also permits single emitter experiments \cite{Albrecht13}. The observed point spread function for individual FNDs is fit by a Gaussian with a $1/e^2$ radius of $w=1.1~\mu$m. This is larger than the value found when the sample faces the objective ($w=0.75~\mu$m) due to the reduced effective numerical aperture, $\mathrm{NA}'\approx \mathrm{NA}/n_m$ when focusing through a dielectric medium with refractive index $n_m$. For time resolved measurements with FNDs on the mirror we find fluorescence lifetimes ranging between 16 and 23~ns. This is on the short side of the typical range for nanodiamonds and can be attributed to a lifetime reduction due to the modified density of states on the mirror.

The observed fluorescence in the measurements shown below corresponds to roughly $50$ NV centers, again estimated from the comparison with single emitter fluorescence yield. With the respective FND positioned in the focus of the microscope objective, we align the fiber to form a cavity such that the fundamental mode has optimal overlap with the FND. Figure~\ref{fig:Results}(c) shows a typical spectrum obtained under such conditions. To understand the different parts of the spectrum, we show a typical free space spectrum taken on a Suprasil substrate with negligible background [Fig.~\ref{fig:Results}(a)] and a calculation of the mirror transmission $T_2(\lambda)$ [Fig.~\ref{fig:Results}(b)]. Multiplying the two contributions and scaling with a constant matches the modulated emission observed under cavity conditions for the short-wavelength part [Fig.~\ref{fig:Results}(c)]. Within the stop band of the coating we observe enhanced emission into sharp cavity resonances. Subsequent fundamental modes at $\lambda_1 = 715~$nm and $\lambda_2 = 755~$nm and a series of weak higher-order transverse modes are visible. The separation of the fundamental modes gives direct access to the effective cavity length $\deff=\lambda_1\lambda_2/[2(\lambda_2-\lambda_1)] = 6.75~\mu$m. This length contains the part of the cavity mode penetrating into the mirror stack. From a simulation of the coating at the respective wavelengths we find that the penetration adds up to $1.5~\mu$m. Evaluating the separation of the higher order transverse modes $\Delta\lambda_{nm}=\lambda^2 \arccos{\left(\sqrt{1-\deff/r}\right)}/(2\pi\deff)$, we infer the effective radius of curvature of the fiber mirror, $r = 100~\mu$m, in agreement with white light interferometric measurements of the fiber surface. This determines the mode waist $w_0$ and volume $V$ of the plane-concave cavity, given by $w_0^2 = \lambda/\pi\sqrt{r\deff-\deff^2}$ and $V=\pi w_0^2\deff/4$, respectively. Since $\deff \ll r$ and $w_0 \gg \lambda/2$, the paraxial approximation and thus the given formulas are valid to a high precision.

To characterize the background contribution, we retract the fiber and perform saturation measurements. We observe an intensity dependence that contains a saturating part and a linear contribution from background fluorescence, shown in Fig.~\ref{fig:Results}(e). Fitting the integrated count rate of the transmissive part to the function $P=P_\infty I/(I_\mathrm{sat}+I)+aI$, with the intensity $I$ and the fitting parameters $P_\infty,I_\mathrm{sat},a$, we find a saturation intensity $I_\mathrm{sat} = 3.5~$GW/m$^2$ for the studied FND, which agrees with typical reported values \cite{Kurtsiefer00}, and a background contribution of $20\%$ in the linear regime of the NV fluorescence. To determine $I$, we measured the excitation mode waist to be $1.5~\mu$m by scanning a FND through the focus and detecting the fluorescence without spatial filtering. Within the mirror stopband and for the cavity present we find a signal to background ratio of about 100 for strong cavity resonances. We correct for the background contributions in the subsequent evaluations. Taken together, the spectrum, saturation, and lifetime measurements confirm that we are indeed coupling NV emission to the cavity.


\begin{figure*}[tb]
\includegraphics[width=0.85\textwidth]{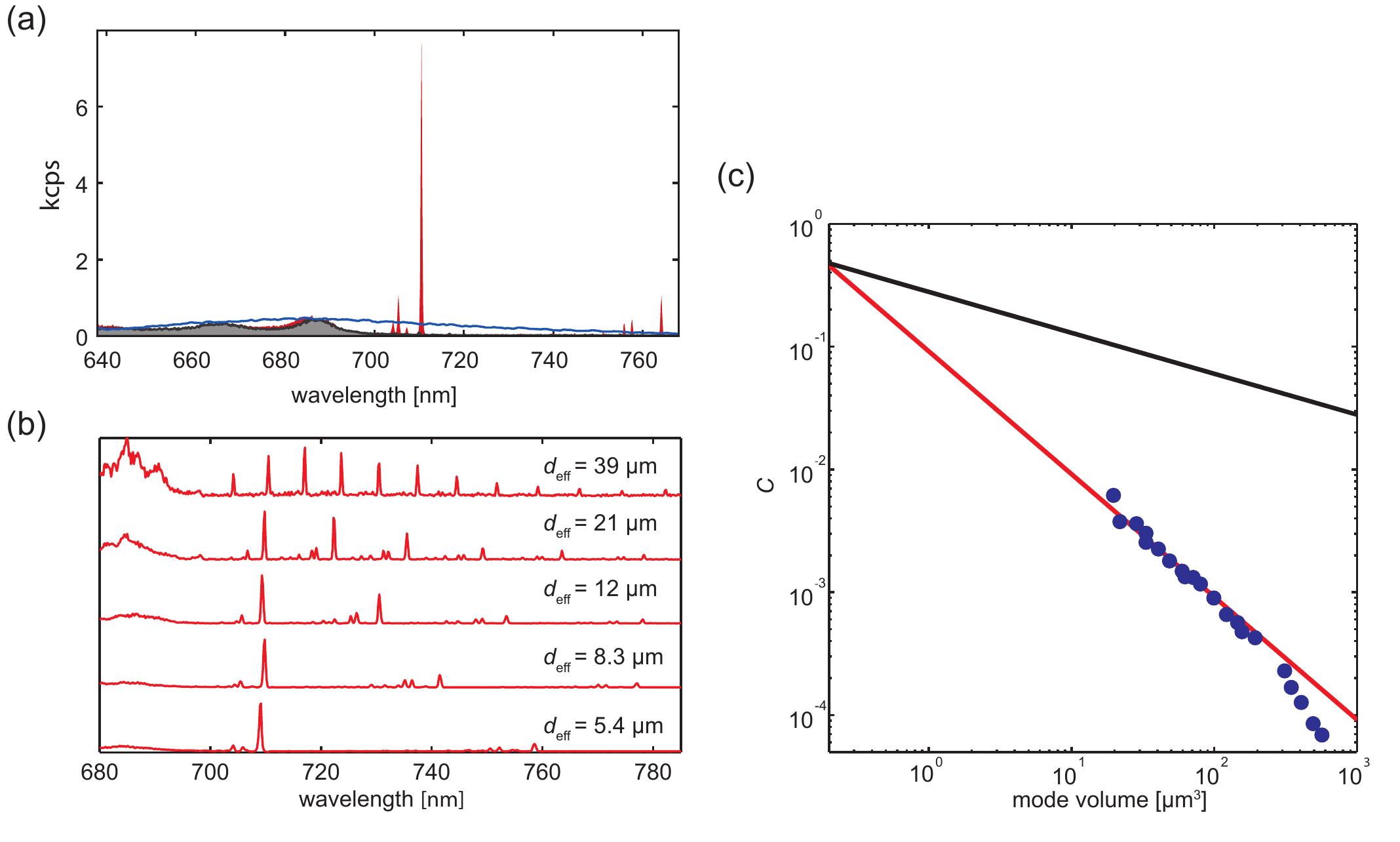}
\caption{\label{fig:Ceff} (Color online) (a) Comparison of the FND spectrum as inferred for free space conditions ($S_{fs}$, blue), emission through the mirror with the fiber retracted ($S_m$, gray), and a cavity configuration with $\deff = 5.0~\mu$m ($S_c$, red). (b) Normalized spectra for decreasing cavity length from $\deff=39~\mu$m (top) to $5.4~\mu$m (bottom). (c) Scaling behavior of the Purcell enhancement in the bad emitter regime: Effective Purcell factor evaluated for the strongest cavity resonance at $\lambda_0 = 710~$nm as a function of mode volume (blue data points). The red line shows the prediction of Eq. (2), and the black line shows the ideal Purcell factor as given by Eq. (1) divided by a factor of 100 for easier comparison. 
}
\end{figure*}

\textit{Effective Purcell Factor.} The ideal Purcell factor $C_0$ describes the enhancement of the emission rate of an emitter whose entire fluorescence spectrum is coupled to an optical cavity resonance with quality factor $Q=\lambda_0/\delta\lambda$ and mode volume $V$,
\be
	 C_0 = \frac{3\lambda^3}{4\pi^2}\frac{Q}{V}\left|\frac{\vec{\mu}\vec{E}}{\mu E_0}\right|^2.
\ee
Here, $\vec{\mu}$ is the dipole matrix element, $\vec{E}$ is the local electric field at the position of the emitter, and $E_0$ is the field maximum in the standing wave antinode.
For the bad emitter regime, i.e., for broadband emitters where electron-phonon interactions distribute the transition over a spectrum $ S(\lambda)$ with spectral width $\delta\lambda_{em}$ much larger than the full cavity linewidth $\delta\lambda$, only a fraction $\delta\lambda / \delta\lambda_{em}$ of the emission couples to the cavity. This reduces the Purcell factor to 
\be
C \approx \eta_\lambda\frac{\delta\lambda}{\delta\lambda_{em}} C_0 = 
\frac{3\lambda_0^3}{4\pi^2}\frac{Q_{em}}{V} \eta_\lambda \left|\frac{\vec{\mu}\vec{E}(\lambda_0)}{\mu E_0(\lambda_0)}\right|^2
\ee
at the cavity resonance wavelength $\lambda_0$, where $\eta_\lambda = S(\lambda_0)/\max(S)$ accounts for a detuning from the maximum of the spectrum.
In this regime, the effective Purcell factor no longer depends on the quality factor of the cavity but rather on the quality factor of the emitter, $Q_{em} = \lambda_0/\delta\lambda_{em}$.

For an experimental evaluation of the effective Purcell factor we compare background corrected spectra for three different configurations: a typical free space reference spectrum $S_0(\lambda)$ from an FND on a Suprasil substrate (sample facing the objective), and spectra $S_c(\lambda)$ and $S_m(\lambda)$ from the FND under study on the mirror (sample facing the fiber) with and without the fiber mirror being present to form a cavity, respectively.
We calculate the expected free space emission spectrum $S_\mathrm{fs}(\lambda)=b S_0(\lambda)$ of the FND under study from its emission spectrum obtained through the mirror by matching $S_m(\lambda)$ to $b T_2(\lambda) S_0(\lambda)$ via the scaling factor $b$ determined by a least square fit. Here, $T_2(\lambda)$ is the transmission of the macroscopic mirror which we infer from a direct transmission measurement and simulate with a transfer matrix method \cite{Hood01}.

With this at hand we can compare the integrated emission into the strongest cavity resonance
$P_c=\int_{\lambda_0-\Delta}^{\lambda_0+\Delta}S_c(\lambda)d\lambda$
with the integrated emission under free space conditions
$P_{fs}=\int S_{fs}(\lambda)d\lambda$,
where $\Delta$ denotes an integration range of 3 full linewidths (FWHM) around the resonance, and the free space spectrum is integrated from 590 to 770~nm to cover the entire emission.
The Purcell factor compares the emission rate into a particular cavity mode with the free space emission into the entire solid angle, such that one has to account for the probability for emitted photons to reach the detector for both cases. For the free space situation in our case, the presence of the Bragg mirror and the substrate modifies the emission pattern. We perform analytical and finite difference time domain simulations to study this aspect. Since we collect without immersion liquid, increased emission into the dielectric beyond the critical angle remains trapped in the medium. Additionally, the angular distribution within the NA of our objective remains nearly unaffected by the refraction through the mirror layers and the substrate in the transmissive spectral range. Consequently, the collection efficiency can be determined from the free space dipole emission pattern, the effective solid angle of the objective as determined from the observed point spread function, and the angle dependent transmission coefficient. Weighting with a dipole orientation distribution (see below) yields a collection efficiency $\eta_{\Omega} = 0.05$.
For a photon emitted into the cavity, we include the probability to leave the cavity through the respective mirror,
$\eta_c=T_2/(T_1+T_2+L_1+L_2)$, where $L_1$ and $L_2$ are the loss of the fiber mirror and the macroscopic mirror respectively. At the wavelength of the cavity resonance, $\lambda_0 = 710~$nm, we obtain $\eta_c = 0.68$, where we use the measured values $L_{1,2} = (71,23)~$ppm, and $T_{1,2}(\lambda_0) = (810,1900)~$ppm from the coating simulation.
Losses on the way to the detector and detector efficiency are assumed to be equal for free space and cavity conditions and thus cancel out.

Overall, this yields an experimental value for the effective Purcell factor
\be
C(\lambda_0) = \frac{P_c(\lambda_0)}{P_{fs}}\frac{\eta_\Omega}{\eta_c}.
\ee
For the spectrum shown in Fig.~3(a) we obtain a value $C = (6 \pm 2) \times 10^{-3}$. The error is estimated from systematic uncertainties of our evaluation method, mainly originating from differences between the spectra of different FNDs and from an alignment and wavelength dependent collection efficiency. Pulsed excitation measurements with and without cavity show no significant difference of the emission lifetime, as expected for the low value of $C$.

To obtain quantitative agreement between the prediction of Eq.~(2) and our measurements evaluated according to Eq.~(3), we estimate the influence of the term $\eta_\lambda|\vec{\mu}\vec{E}(\lambda_0)/(\mu E_0(\lambda_0))|^2$. Over the studied spectral range, the position of the standing wave maximum shifts from the position of the emitters at the center wavelengh towards the coating for shorter wavelengths. This reduces the coupling in particular close to the edge of the stop band of the mirror, where a phase shift results from the dispersive character of the coating. For the resonance position ($\lambda_0 = 710~$nm) and our coating parameters we calculate $\eta_E = (E/E_0)^2 = 0.55$. 
Next, the detuning from the maximum of the NV spectrum reduces the coupling by a factor $\eta_\lambda= 0.75$, which we read off from the reference spectrum.
Finally, we estimate the effect due to the fixed but random dipole orientations. The measurements were performed with circular polarization and excitation conditions below saturation, such that the angle dependence of the excitation probability $(\vec{\mu}\vec{E})^2\propto \cos^2(\theta)$ leads to a normalized distribution $p(\theta) = 3/2\cos^2(\theta)$ with $\theta$ being the angle between the mirror plane and the dipole axis. This reduces the observed average enhancement by a factor $\eta_\theta = \int p(\theta)\cos^2(\theta)\cos(\theta) d\theta = 0.8$.
Evaluating Eq.~(2) for the parameters of the measurement ($V = 19~\mu$m$^3, Q_{em}\approx 10$) we obtain 
$C = 3\lambda_0^3/(4\pi^2)Q_{em}/V\times \eta_\theta \eta_\lambda \eta_E= (5 \pm 2)\times 10^{-3}$, in good agreement with the experimental value.

To demonstrate the scaling of Eq.~(2) explicitly, we perform measurements as shown in Fig.~\ref{fig:Ceff}(a) and evaluate the effective Purcell factor for the emission into the strongest cavity resonance as a function of the cavity mode volume by changing the mirror separation. A few example spectra are shown in Figure \ref{fig:Ceff}(b), and the evaluation of the full data set is shown in (c). 
The observed behavior matches the prediction of Eq.~(2) very well. 
Notably, $C$ approaches unity only for very small mode volumes on the order of $\lambda^3$. We expect that this regime can be approached with fiber-based Fabry-Perot cavities by further miniaturization.

\begin{figure*}[ht!]
\includegraphics[width=0.85\textwidth]{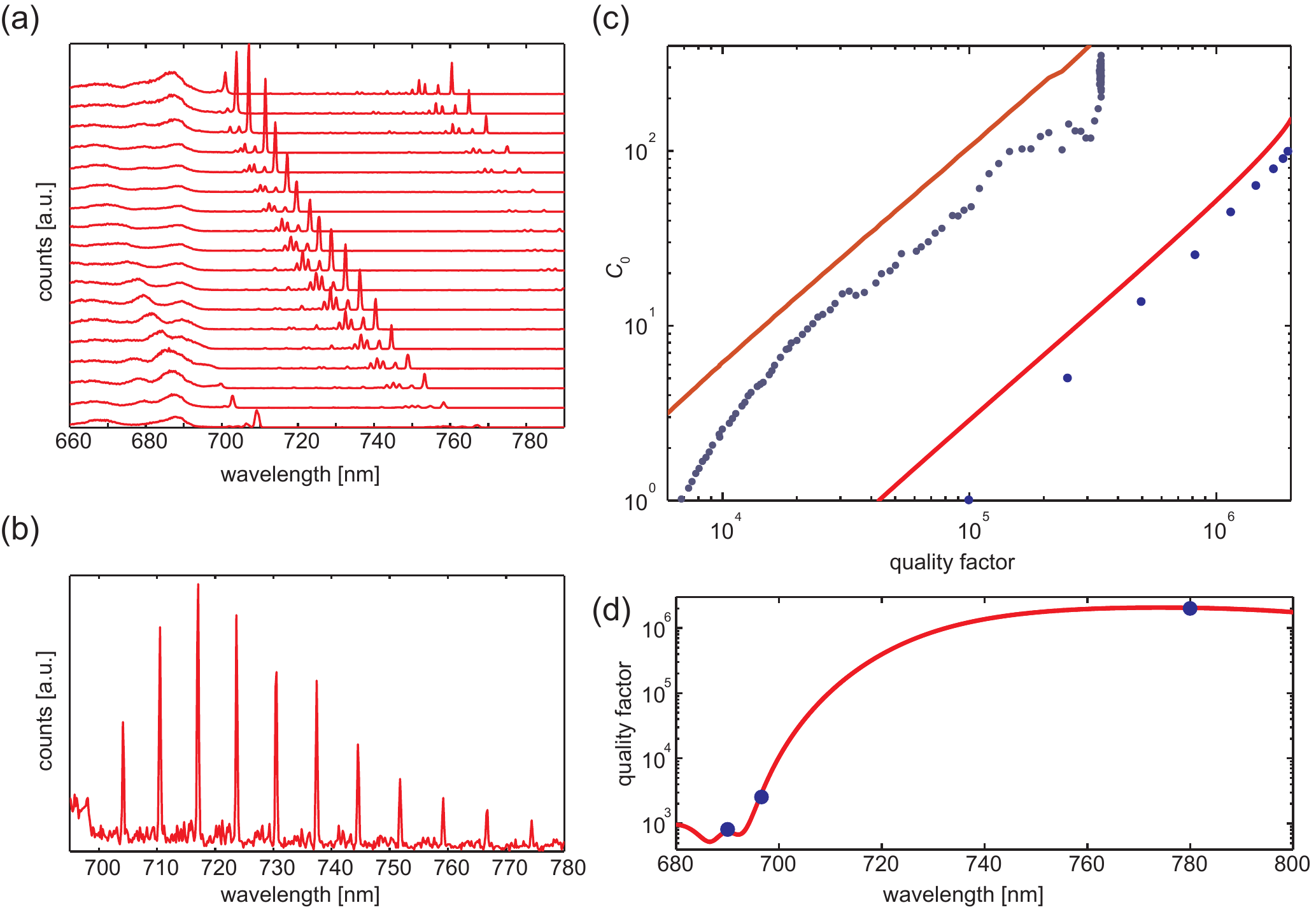}
\caption{\label{fig:Cideal} (Color online) (a) Tuning of a single cavity resonance across the accessible NV spectrum starting at $\deff = 4.3~\mu$m. (b) Cavity spectrum with multiple resonances for $\deff = 39~\mu$m. (c) Ideal Purcell factor as a function of the quality factor for the data set partially shown in (a) (light blue) and (b) (dark blue) together with the predictions of Eq.~(1) (solid lines). (d) Measurement (blue datapoints) and calculation (solid line) of the cavity quality factor as a function of the wavelength for $\deff=39~\mu$m.
}
\end{figure*}

\textit{Ideal Purcell Factor.} In contrast to the small fraction of the entire emission that is coupled to the resonator, we observe a strong enhancement of the emission spectral density on cavity resonance. In fact, comparing the spectral density for the cavity and free space case quantifies the ideal Purcell factor $C_0$. This can be seen by considering a spectral element of the emission with Fourier limited width, which couples to the cavity resonantly and thereby experiences the full enhancement.
With the strong variation of our coating properties across the emission spectrum, we can study the enhancement as a function of the quality factor of the cavity. By changing the mirror separation we can locate the cavity resonance at any wavelength and thereby sample quality factors spanning a range of almost three orders of magnitude. Figure \ref{fig:Cideal}(a) shows the stepwise tuning of a cavity resonance across one free spectral range for a short cavity corresponding to the mode with eight antinodes between the mirrors (smallest $\deff = 4.3~\mu$m). For comparison, we show a single spectrum for large cavity length $\deff = 39~\mu$m in Fig.~\ref{fig:Cideal}(b), where the cavity geometry and the excitation conditions are fixed. There, several consecutive fundamental modes between 700 and 800~nm are observed within the mirror stop band, sampling different $Q$ within a single measurement. 

The ideal Purcell factor can be evaluated by comparing the maximal spectral density on resonance $S_{c,\mathrm{max}}(\lambda_0)$ with the free space value $S_{fs}(\lambda_0)$ at the same wavelength. Since the cavity linewidth is beyond the resolution of our spectrometer, we 
infer the peak spectral density of a Lorentzian resonance
$S_{c,\mathrm{max}} =2P_c/\pi\delta\lambda$ 
and use 
the cavity linewidth
$\delta\lambda = \lambda_0^2(T_1+T_2+L_1+L_2)/4\pi d\sqrt{R_1R_2}$,
where $R_i = 1 - T_i - L_i$. For $\lambda_0 = 780~$nm we measure $Q=\lambda_0/\delta\lambda$ with a narrowband diode laser, while for low $Q$ we use broadband light and the spectrometer.
A calculation of $Q(\lambda)$ together with the measurements are shown in Fig.~\ref{fig:Cideal}(d).
The free space spectrum as well as corrections due to the collection efficiency and mirror loss are treated in analogy to the previous evaluation. Together, the ideal Purcell factor is determined experimentally by 
\be
C_0 = \frac{S_{c,\mathrm{max}}}{S_{fs}}\frac{\eta_\Omega}{\eta_c}.
\ee
In Fig.~\ref{fig:Cideal}(c) the values obtained from the data sets partially shown in Fig.~\ref{fig:Cideal}(a),(b) are compared with theory, where we again include $\eta_E$ and $\eta_\theta$. Since $S_{c,\mathrm{max}}$ and $S_{fs}$ are evaluated at the same wavelength, the same detuning factor applies for both and thus drops out. For $\deff = 4.3~\mu$m and $Q = (3.5\pm 0.3)\times 10^5$, the measurements yield a maximum value of $C_0 = 300 \pm 100$, one of the largest values reported to date. 
Notably, we observe only minor changes of the finesse (or the quality factor) measured at the largest value when introducing the nanocrystal ($< 20\%$). 

\textit{Rate model.} In the previous sections we have approximated the NV emission by a single transition with large effective linewidth to simplify the discussion. A more accurate treatment accounts for the individual transitions making up the emission spectrum and the respective rates involved. The coupled system can therefore be described in the framework of the dissipative Jaynes Cummings theory including excessive dephasing \cite{Auffeves10,Valle11}. We consider the dynamics governed by the coherent coupling rate $g_0$, the cavity decay rate $\kappa$, the decay rate of the emitter $\gamma_0$, and the excessive dephasing rate $\gamma^*$. Coupling to phonons distributes the dipole matrix element between the ZPL and individual sideband transitions. To simplify the description of the sideband, a single phonon mode picture is used, where the $k$-th phonon sideband is approximated by a Lorentzian with effective width $\gamma_k^*$ and transition strength $\zeta_k$. In this way we can treat the ZPL ($k=0$) and the sideband ($k=1 \dots n$) in the same manner. For an ensemble, inhomogeneous broadening can contribute an additional dephasing mechanism, which, however, is expected to be small compared to the phonon induced broadening in the sideband. The coupling rate for a single emitter and an individual transition is then given by
\be
g_{0k}=\frac{\vec{\mu_k}\vec{E_k}}{\hbar} = \sqrt{\frac{3\pi c^3}{2\omega_k^2V}\zeta_k\eta_E\gamma}\cos{\theta}.
\ee
For the ensemble, we replace $\cos{\theta}$ with $\sqrt{\eta_\theta}$ to account for the orientation distribution.
Finally, the energy decay rate of the cavity at the respective transition frequency is given by $\kappa_k = c(T_1+T_2+L_1+L_2)/2\deff\sqrt{R_1R_2}$ and $\gamma = 1/\tau = 2\pi \times 8~$MHz for typical nanodiamonds.

Generalizing the results from \cite{Auffeves10,Valle11} one finds that the emission rate of an emitter into a single cavity mode at frequency $\omega_c$ is given by
\be
R(\omega_c) = \sum_k\frac{4g_{0k}^2\Gamma_k}{\Gamma_k^2+4\Delta_k^2}.
\ee
Here we account for all contributing transitions with respective frequencies $\omega_k$, coupling strengths $g_{0k}$, total incoherent rates $\Gamma_k = \kappa_k+\gamma_0+\gamma_k^*$, and cavity detunings $\Delta_k = \omega_c-\omega_k$. The effective Purcell factor is then given by $C(\omega_c) = R(\omega_c)/\gamma_0$. One can see that $C$ becomes independent of $\kappa_k$ for $\gamma_k^*\gg \kappa_k$, supporting the previous definition in Eq.~(2).

We note that the coupling of a number of $N$ emitters to the cavity does not lead to a change of the coupling rate in the present situation. This is in contrast to the case of a coherent ensemble, where a collective dipole forms that leads to an increased coupling rate $g_N = \sqrt{N}g_0$. Collective enhancement in the presence of strong dephasing can be quantified by the figure of merit $\alpha = NR/\gamma^*$, which compares the enhanced emission rate under ideal conditions with the dephasing rate \cite{Temnov05}. While modification of the dynamics becomes apparent for $\alpha > 0.1$, we find $\alpha \sim 10^{-3}$ for our parameters, such that no collective effects are expected.

To model the measurements of our experiment, we fit $\zeta_k$ and $\gamma_k*$ with $\omega_k = 2\pi (470 - k\times16)~$THz to reproduce the free space spectrum $S_0(\omega)$, and determine all other quantities from measurements. We find $\zeta_k = [0.02, 0.25, 0.44, 0.24, 0.06, 0.01], \gamma_k^* = 2\pi [3, 23, 25, 29, 34, 40]~$THz, contributions of transitions with larger $k$ are negligible. The large difference in $\gamma_k^*$ for the ZPL and the sidebands reflects their slightly different origin. While for the ZPL, only thermally excited low frequency phonons contribute, each sideband is made up by the coupling to a structured phonon continuum extending up to the cutoff at $\sim 40$THz \cite{Davies74}. Furthermore, the width of the bands increases with $k$ due to the increasing number of possible phonon decay paths.
We find good agreement between the predictions of the model and the data, e.g. for the parameters of the measurement shown in Fig.~\ref{fig:Ceff}(a) we obtain $C = (6\pm 2)\times 10^{-3}$.
The obtained transition strengths deviate from the expected Frank-Condon factors, which in part could be attributed to a wavelength dependent collection efficiency in our setup. This would imply an error in the evaluated Purcell factors of up to $\sim 20\%$ in both models and the experimental data.

The generalized description of cavity enhancement confirms that the simple picture given previously is a sufficient approximation as long as the effective dephasing dominates all other rates and masks the structure of the phonon sideband. Beyond this, the model provides insight into the different roles of transition branching and dephasing.

\textit{Conclusion and Outlook.} In our experiment we have studied the scaling laws for the cavity enhancement of the fluorescence of broadband emitters with a flexible architecture, offering high finesse and large ideal Purcell factors compatible with the presence of nanodiamonds. The results indicate a method for the realization of efficient single-photon sources with narrow bandwidth and wide tunability, and suggest that fiber-based microcavities could provide a route towards the strong coupling regime in a cryogenic environment. 

While NV centers in nanocrystals show larger homogeneous linewidth and stronger spectral diffusion than in clean bulk samples, dephasing rates $\gamma^*/2\pi< 500~$MHz for the ZPL appear feasible. Alternatively, diamond membranes or nanomachined high quality bulk diamond could be introduced to achieve better optical and spin properties. 
For an estimation of the achievable coupling strength we account for the ZPL branching ratio $\zeta_0= 0.04$ and the splitting of the ZPL into six individual fine structure resonances. For the transitions connecting the $m_S = 0$ ground state with the $E_x$ or $E_y$ levels in the excited state, an overall branching ratio $\zeta_0(E_x,E_y) \approx 0.02$ remains for perfect state initialization \cite{Manson06}. Optimization of the laser machining technique is expected to achieve mirror radii of curvature $<10~\mu$m, which would result in $2g_{00}>2\pi\times1~$GHz for $\deff=2.0~\mu$m. 
Applying a feasible coating with $T +L = 20~$ppm at 637~nm yields $\F=150000$ and $\kappa/2\pi  = 500$ MHz at this mirror separation. With a penetration depth $\sim 0.6~\mu$m at the coating center, the mirror separation still amounts to $\sim 2\lambda$, providing sufficient space for alignment and tunability. Together, the assumed parameters would reach the strong coupling regime, $2g_{00} > (\kappa,\gamma,\gamma^*)$, and lead to an effective Purcell Factor $C > 130$.
This opens the perspective for a fully quantum-coherent spin-photon interface.

We acknowledge support from L. Costa, M. Mader, M. Weber, L. Liebermeister, and D. Burchardt, and helpful discussions with  R. Albrecht, C. Becher, and A. Kubanek.
This work was supported by the Nanosystems Initiative Munich. T.W.H.\ gratefully acknowledges support by the Max Planck Foundation.




\end{document}